\documentclass[10pt]{iopart}

\usepackage{harvard}
\usepackage{graphicx}
\usepackage{undertilde}
\usepackage{wrapfig}

\usepackage{xcolor}

\usepackage{iopams}

\citationmode{abbr}
 
\begin{document}

\title{Modeling of an electro-active pseudo-trilayer based on PEDOT, a semi-conductor polymer}

\author{M. Tixier$^1*$ \& J. Pouget$^2$}

\address{$^1$ Université Paris-Saclay, UVSQ, CNRS, Laboratoire de Math\'ematiques de Versailles (LMV), UMR 8100, 45, avenue des Etats-Unis, F-78035 Versailles, France}
\address{$^2$ Sorbonne Universit\'e, CNRS, Institut Jean le Rond d'Alembert, UMR 7190, F-75005 Paris, France}
\ead{\mailto{mireille.tixier@uvsq.fr}, \mailto{joel.pouget@upmc.fr}}
\vspace{10pt}
\begin{indented}
\item * Author to whom any correspondence should be addressed
\vspace{10pt}
\item[]August 2024
\end{indented}

\begin{abstract}
Electroactive polymers are smart materials that can be used as actuators, sensors, or energy harvesters. We focus on a pseudo trilayer based on PEDOT, a semiconductor polymer : the central part consists of two interpenetrating polymers and PEDOT is polymerized on each side ; the whole blade is saturated with an ionic liquid. A pseudo trilayer is obtained, the two outer layers acting as electrodes. When an electric field is applied, the cations move towards the negative electrode, making it swell, while the volume decreases on the opposite side ; this results in the bending of the strip. Conversely, the film deflection generates an electric potential difference between the electrodes.
We model this system and establish its constitutive relations using the thermodynamics of irreversible processes ; we obtain a Kelvin-Voigt stress-strain relation and generalized Fourier’s and Darcy’s laws. We validate our model in the static case : we apply the latter to a cantilever blade subject to a continuous potential electric difference at the constant temperature. We draw the profiles of the different quantities and evaluate the tip displacement and the blocking force. Our results agree with the experimental data published in the literature.
\end{abstract}

\vspace{2pc}
\noindent{\it Keywords }: Electro-active polymer (EAP), PEDOT, Ionic polymer modeling, Polymer mechanics, Multiphysics coupling, Constitutive relation, Smart material
\newline
\newline
published in \textit{Smart Materials and Structures} \textbf{33}(095030), 2024

https://doi.org/10.1088/1361-665X/ad6ab8

\ioptwocol

\section{Introduction}

Electro-active polymers (EAP) are materials with interesting unusual properties that can be used for designing and constructing structures, 
and they are referred to as smart materials. More specifically, electro-active polymers are materials whose the shape is modified when an electric field is 
applied to them. They also undergo structural deformations such as swelling, shrinkage, or bending in response to an electric stimulus. 
As a result, they can be used as actuators or sensors. Actuators designed with EAPs produce a large amount of deformation while sustaining 
significant forces. These properties make these materials suitable for use in advanced engineering devices, including micro-robotics 
\cite{sha1994,Morton2023}. The growing interest in smart materials has driven for scientists to seek inspiration from  
living systems in the design of materials that mimic the behavior of living organisms  \cite{Zhao,Shen,Aureli2009,Chen2017}. Therefore, 
it seems logical to take advantage of the properties of the active/reactive polymers such as EAP in medical technology \cite{Fattah2024,Zhang2021}. 
Notably, EAPs find interesting application in  the medical endoscope \cite{jon2007}, bio-sensors, chemico-mechanical actuators \cite{Olvera2021}, and artificial muscles 
\cite{bar-cohen2005,Wang2023,Chen2023}. The biocompatibility of the IPMC is advantageous for the use in biosensor devices capable of measuring  
human body activity \cite{chikhaoui2018}. EAPs find their most interesting applications in tunable and adaptative  devices across a broad range of industrial 
domains \cite{Pelrine,OHalloran}. \\
 
\noindent 
Depending on the electro-mechanical activation mechanism EAPs can be categorized into two main types : electronic electro-active polymers 
(EEAPs) and ionic EAPs. In the first category, two electro-mechanical effects are mainly responsible for the deformation of the polymer : (i) the Maxwell strain  
\cite{Maugin} due to Coulomb forces acting within the material , and (ii) the electrostriction, which is caused by the intermolecular electrostatic forces \cite{ask2012}. \\

\noindent 
The movement of ions within the EAPs is caused by an electric field generated by an electric potential difference applied to the metallic 
electrodes (usually made of gold or platinum) located on the upper and lower faces of the polymer membrane.  This composite material is known as an ionic-polymer 
material composite (IPMC).  The bending deformation of an IPMC is used in various actuators 
applications, such as grippers used to pick up small objects  \cite{deole2008,Ford,Vogel}. By changing the direction of the electric field the bending direction is also reversed. 
This property is used for fabricating diaphragms for volumetric pumps \cite{Schomburg}. The EAPs find their applications in bio-engineering, such as  designing 
tactile displays for visually impaired people \cite{Chouvardas,Vitushinsky,Feng}.  EAPs are also used in field of soft robotics \cite{Rohtlaid2021}, where a 
kind of caterpillar model made of sections of EAP actuators can be imagined, and in smart personal protective equipment  \cite{Zhang2021,Dutta2022}. Extension to spatial 
applications has been reported in \citeasnoun{FannirTh}. Sensor process can occur in reverse, meaning  if the EAP undergoes a deformation or if a 
force is applied to the polymer, an electric potentiel difference can be measured on the electrodes \cite{MohdIsa}  caused by the change in the ion concentration in 
the polymer blade \cite{Bonomo,Dominik}. EAP thin stripe can be used as bending sensor making the polymer promising candidate for energy harvesting \cite{brufau-penella2008,Cellini2014a,Cellini2014b}.  \\
 
\noindent 
In conclusion, despite the relatively slow response time and the moderate actuator force, ionic EAPs remain advantageous due to their 
low activation voltage (a few volts) and large bending displacements. This smart material is becoming an increasingly attractive source of inspiration for researchers 
and engineers.  \\
 
\noindent 
A key focus of our current research is to develop a continuous medium model for electroactive polymers of the ionic class. 
Our approach, adopted from multi-phase problems enables us to transition from a microscopic description of the 
material to a macroscopic one. We establish the conservation equations at the microscopic scale for each phase and the interfaces. The macroscopic 
equations for the polymer are then derived by averaging the corresponding microscopic quantities weighted by a function of presence. This includes the balance 
equations for mass, momentum, total, kinetic, potential and internal energy densities, entropy as well as the balance equation for 
electric charge, and the Maxwell equations. Using the thermodynamics of linear irreversible processes we can deduce the constitutive equations 
\cite{deGroot,Tixier1,Tixier2}. In this study, we focuse on modeling a thin strip of Nafion, and we compare the results  to experimental 
data available  in the literature \cite{Tixier4},  successfully validating the proposed model. However, there remains a significant challenge 
to enhance the performances of  electro-active polymers both for both actuator and sensor functions. We aim to apply this thermodynamic 
approach to a conducting polymer thin trilayer stripe based on an interpenetrated polymer network (IPN) \cite{festin2014,Festin2013}. This alternative to 
traditional electro-active polymers, such as those based on Nafion or other polyelectrolytes like Flemion or Aciplex.{has shown promising  
performance enhancements in recent studies \cite{Seurre2023,Catry,Rohtlaid2021}. This new class of EAP is cpable of functioning in open-air and even in vacuum, 
providing significant mechanical amplifications and demonstrating a notable time response dynamic stimuli  \cite{Hik2023} (few hundred of Hertz)  
while also preventing and there is no electrode delamination \cite{nguyen2018}. \\

\noindent
The paper is structured as follows. Section 2  describes and models the trilayer electro-active polymer. Section 3 reports the micro-macro approach 
of the polymer, along with the conservation laws and the constitutive equations. Section 4 applies the model to the bending of a clamped-free beam under an 
electric potential difference applied to the lower and upper active layers. It also computes the tip displacement and the blocking force as functions of the 
material parameters in the static and isothermal cases. Comparison with experimental results found in the literature are reported and discussed in Section 5.
Section 6 draws the conclusions.

\section{Description and modeling of the material}

The material being studied was developed by a team of chemists \cite{Festin2013,festin2014}. It consists of a pseudo trilayer composed of three interpenetrated polymers soaked in an ionic liquid. Interpenetrated means that the polymers cannot be separated without breaking the covalent bonds, which prevents the delamination of the electrodes. The first polymer, PEO (polyethylene oxide) is chosen for ionic conduction. The second polymer, NBR (acrylonitrile-butadiene copolymer) polymerizes within the matrix formed by the first polymer. NBR is an elastomer that improves the mechanical properties of the mixture. The NBR used in the studied pseudo trilayers contains $44\%$ 
acrylonitrile \cite{Festin}.

The film is composed of $60~wt \%$ PEO and $40~wt \%$ NBR. Both sides of the film are then impregnated with the precursor of a semiconductor polymer, PEDOT (poly (3, 4 - ethylenedioxythiophene)), which is mainly concentrated near the two sides. The mixture is saturated with an ionic liquid, EMITFSI (1-ethyl-3-methyl-imidazolium bis (trifluoromethanesulfonyl) imide, \fref{EMITFSI}), which penetrates almost exclusively into the central part of the film. The penetration of the EMITFSI is facilited by the PEO, which creates free volume due to its pending chains, and its polarity allows the dissociation of the ionic liquid \cite{Das,Karmakar}. These four material components 
together form a pseudo trilayer $250\mu m$ thick, with the two outer layers rich in PEDOT acting as electrodes and the central part as an ion reservoir. The composition is 
optimized for the intended applications : $13.3~wt \%$ NBR, $20.0~wt \%$ PEO, $9.4~wt \%$ PEDOT and $57.3~wt\%$ EMITFSI \cite{festin2014}.

\begin{figure}[h]
\includegraphics[height=4.5cm]{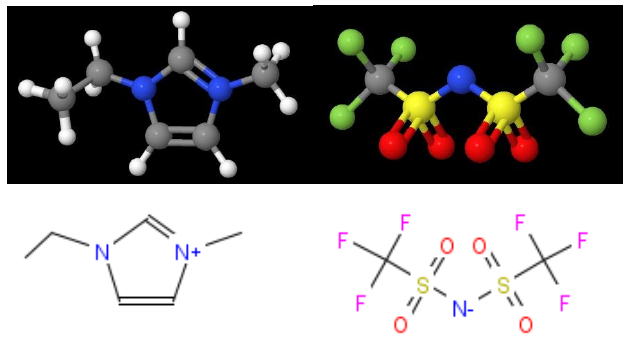}
\caption{Chemical formulae for the ions $EMI^{+}$ and $TFSI^{-}$.}
\label{EMITFSI}
\end{figure}

\begin{figure*}[h!]
\includegraphics[width=1.0\textwidth]{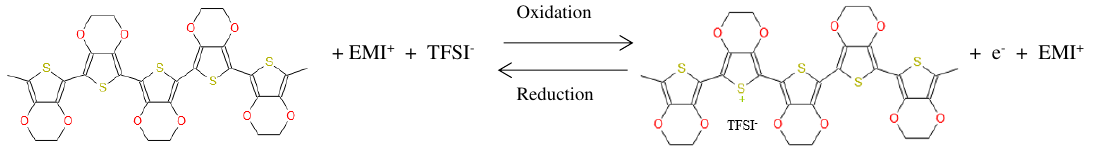}
\caption{Oxidation/reduction reactions of PEDOT.}
\label{Reactions-PEDOT}
\end{figure*}

The EMITFSI consists of $EMI^{+}$ cations, which can move within the layer, and $TFSI^{-}$ anions, which are slightly larger and have very low mobility \cite{Randriamahazaka2004}. When the saturated trilayer is placed in an electric field perpendicular to its faces, the PEDOT undergoes an oxidation reaction (or p-type doping) in the positive electrode (\fref{Reactions-PEDOT}). This causes an inflow of cations $EMI^{+}$ from the central part and an increase in volume on the opposite side, resulting 
in the blade bending towards the positive electrode. The $TFSI^{-}$ anions remain embedded in the polymer network (\fref{Meca-flexion}).

\begin{figure}[h]
\includegraphics[height=3.2cm]{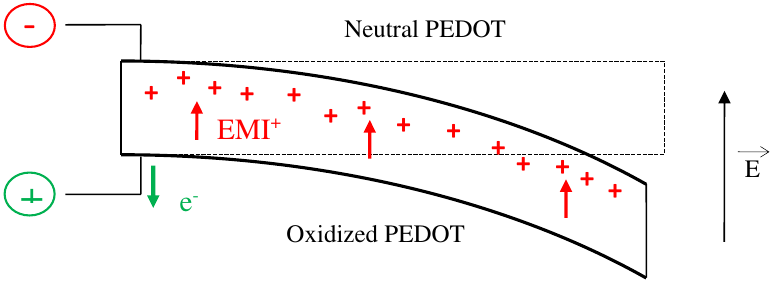}
\caption{Bending mechanism of the pseudo trilayer.}
\label{Meca-flexion}
\end{figure}

The distribution of PEDOT in the thickness of the dry film (i.e. without EMITFSI) can be measured by SEM-EDX spectroscopy \cite{Festin2013}. PEDOT is mainly concentrated over a thickness of $30\mu m$ near the film faces with an average concentration of around $0.40g~cm^{-3}$. It only penetrates slightly into the central part where its concentration is  about  $0.08g~cm^{-3}$. On the other hand, the electrodes absorb very little ionic liquid. This distribution determines several parameters, including the density, the mass charge, 
and the Young's modulus. Throughout the rest of this work we will model them using rectangular functions.

This system, similar to Ionic Polymer-Metal Composites based on Nafion \cite{Tixier3}, can be modeled using the thermodynamics of linear irreversible processes through a two-phase "continuous medium" approach. In this model, the three polymers (PEO, NBR and PEDOT) with the $TFSI^{-}$ anions incorporated into their chains are treated as a deformable, homogeneous, and isotropic solid porous medium moving with a velocity field $\vec{V_{2}}$ ; the $EMI^{+}$ cations form a liquid phase that move at the velocity $\vec{V_{1}}$ 
within the pores. The two phases are separated by an interface without thickness. Additionally, the assumption is made that the liquid phase is incompressible and 
 the deformations of the solid are small. Gravity and magnetic induction are supposed to be negligible.

\section{Model for a PEDOT-based pseudo trilayer}

We employe a similar approach to the one used for the Nafion and other similar materials \cite{Tixier1,Tixier2,Tixier3}. In the follows discussion, 
the subscripts $1$ and $2$ will respectively denote the liquid and solid phases. We utilized a coarse-grained model specifically developed for two-component 
mixtures \cite{Ishii06}.

\subsection{Average process}
To start, the conservation equations are formulated  for each phase and for the interfaces at a microscopic scale. PEO and NBR occupy domains around $200 nm$ in size \cite{Festin2013}, while PEDOT forms micrometer-sized grains \cite{FannirTh}. Typically, the microscopic scale is of the order of $100\AA$. Let $g_{k}^{0}$ be the volume density of an extensive quantity related to phase $k$ ; the superscript $^{0}$ indicates that this quantity is defined on the micro-scale. Its balance equation can be written as 
\begin{eqnarray}
\frac{\partial g_{k}^{0}}{\partial t}+div\left( g_{k}^{0}\vec{V_{k}^{0}}\right) =-div\vec{J_{k}^{0}}+\phi _{k}^{0} ,
\end{eqnarray}%
where $\vec{J_{k}^{0}}$ is the flux of $g_{k}^{0}$ due to phenomena other than convection, and $\phi_{k}^{0}$ its volume production (source term). Subsequently, 
these equations are then averaged at a macroscopic scale, in the order of $10 \mu m$ for the complete material, using a presence function $\chi_{k}$ for each phase and interface 
\begin{equation}
\chi _{k}= \cases{1&when phase $k$ occupies the point\\
0&otherwise\\}
\end{equation}

The average value, symbolically noted by $\left\langle \right\rangle$,  over a volume known as  the Representative Elementary Volume (R.E.V.) is assumed to be equal to a 
statistical average (or expected value) due to an ergodic hypothesis. Additionally, we assume that velocity fluctuations are negligible on the scale of the R.E.V.
\begin{eqnarray}
\frac{\partial g_{k}}{\partial t}+div\left( g_{k}\vec{V_{k}}\right) =-div\vec{J_{k}}+\phi _{k}-\left\langle \vec{J_{k}^{0}}.\vec{n_{k}}\chi _{i}\right\rangle ,
\end{eqnarray}%
where%
\begin{eqnarray}
g_{k}=\left\langle \chi _{k}g_{k}^{0}\right\rangle \quad\vec{J_{k}}=\left\langle \chi _{k}\vec{J_{k}^{0}}\right\rangle  \quad \phi _{k}=\left\langle \chi _{k}\phi _{k}^{0}\right\rangle .
\end{eqnarray}
The subscript $_{i}$ represents the interface and $\vec{n_{k}}$ the outward-pointing unit normal to the interface of the phase $k$.
The balance equations need to be written for a closed system in the thermodynamic sense, i.e. that does not exchange mass. We can define particle derivatives $\frac{d_{k}}{dt},~k=1,2$ or derivatives following the motion of the components 1 or 2. Since the velocity fields of the two components are different, we have introduced a "material derivative" $\frac{D}{Dt}$ which enables us to track each of the components in their own motion : the weighted average of the particle derivatives \cite{Biot77,Coussy95,Tixier1} 
\begin{eqnarray}
\rho \frac{D}{Dt}\left( \frac{g}{\rho }\right) 
&=\sum\limits_{1,2,i} \rho _{k}\frac{d_{k}}{dt}\left( \frac{g_{k}}{\rho _{k}}\right) \nonumber \\
&=\sum\limits_{1,2,i}\frac{\partial g_{k}}{\partial t}+div\left( g_{k} \vec{V_{k}} \right) ,            
\end{eqnarray}
or for a vectorial quantity 
\begin{eqnarray}
\rho \frac{D}{Dt} \left( \frac{\vec{g}}{\rho }\right) %
=\sum\limits_{1,2,i}\frac{\partial \vec{g_{k}}}{\partial t}+div\left( \vec{g_{k}}\otimes \vec{V_{k}}\right) ,
\end{eqnarray}
where $\rho$ represents the density and $\rho_{k}$ defines the mass concentration of phase $k$. By summing the macroscopic equations for the phases and interfaces, 
we deduce the balance equation for the complete material 
\begin{eqnarray}
\rho \frac{D}{Dt} \left( \frac{g}{\rho}\right) =-div%
\vec{J}+\phi ,
\end{eqnarray}%
where the quantities without subscripts are relative to the whole material%
\begin{eqnarray}
\vec{J}=\sum\limits_{1,2,i}\vec{J_{k}} \qquad \qquad \phi =\sum\limits_{1,2,i}\phi _{k} .
\end{eqnarray}

\subsection{Conservation laws}
\medskip
We thus obtain all the balance equations of the system : the mass, the linear momentum and the electric charge conservations, and the Maxwell's equations, 
they read as 
\begin{eqnarray}
\frac{\partial \rho }{\partial t}+div\left( \rho \vec{V}%
\right) =\frac{d\rho }{dt}+\rho div\vec{V}=0 ,\\
\rho \frac{D\vec{V}}{Dt}=div\utilde{\sigma }+\rho Z\vec{E} ,\\
div\vec{I}+\frac{\partial \left( \rho Z\right) }{\partial t}=0 ,\\
rot\vec{E}=\vec{0} ,\\
div\vec{D}=\rho Z ,
\end{eqnarray}
and the constitutive relation, which is that of an isotropic linear dielectric, assuming that phases $1$ and $2$ are similar to isotropic linear dielectrics 
\begin{eqnarray}
\vec{D}=\varepsilon \vec{E} .
\end{eqnarray}
$\utilde{\sigma}$ is the stress tensor, $\vec{E}$ the electric field, $\vec{D}$ the electric induction, $Z$ the mass electric charge, $\vec{I}$ the current density vector and $\varepsilon$ the absolute permittivity. We check that the stress tensor is symmetrical.

\renewcommand{\arraystretch}{2}
\begin{table*}[t!]
\caption{\label{tableE}Energies balance equations.} 
\begin{indented}
\lineup
\item[]\begin{tabular}{lllll}
\br
& \qquad Fluxes &$E_{c}\longleftrightarrow E_{p}$ & $U\longleftrightarrow E_{p}$ & $E_{c}\longleftrightarrow U$ \\ \mr
$\rho \frac{D}{Dt}\left( \frac{E_{p}}{\rho }\right) $&$=$& $- \rho Z \vec{V}.\vec{E}$ & $-%
\vec{i}.\vec{E}$ &  \\
$\rho \frac{D}{Dt}\left( \frac{E_{c}}{\rho }\right)$&$=-div\left(-\utilde{\sigma}.\vec{V} \right)$& $+\rho Z \vec{V}.\vec{E}$ &  & $-\utilde{\sigma}:grad\vec{V}$ \\
$\rho \frac{D}{Dt}\left( \frac{U}{\rho }\right) $&$=-div\left[
\vec{Q'} - \sum\limits_{k=1,2} \utilde{\sigma_{k}}.(\vec{V_{k}}-\vec{V}) \right]$ & & $+\vec{i}.\vec{E}$ & $+ \utilde{\sigma}:grad\vec{V}$ \\
$\rho \frac{D}{Dt}\left( \frac{E_{tot}}{\rho }\right)$&$ =-div\left[ \vec{Q'} - \sum\limits_{k=1,2}\utilde{\sigma _{k}}.(\vec{V_{k}} - \vec{V}) -\utilde{\sigma}.\vec{V} \right]$ & & &\\ \br
\end{tabular}
\end{indented}
\end{table*}
We can also write the balance equations for the kinetic energy $E_{c}=\frac{1}{2} \rho V^{2} \simeq \sum\limits_{k=1,2} \frac{1}{2} \rho_{k} V_{k}^{2}$, the potential energy $E_{p}=\frac{1}{2} \vec{E}.\vec{D}$, the internal energy $U$ and the total energy $E_{tot}=E_{c}+E_{p}+U$ (as shown in \tref{tableE}). It is worthwhile mentioning that the kinetic energy is approximately equal to the sum of the kinetic energies of the components when the relative velocities are neglected. In the relations reported in Table 1, $\vec{Q'}$ represents  
the conduction heat flux and $\vec{i}$ is the diffusion current, i.e. the electric current measured in the barycentric reference frame.

These equations describe the fluxes of the different forms of energy : the internal energy flux is due to the heat conduction and to the work of the contact forces in the barycentric reference frame, the kinetic energy flux due to the work of contact forces in the laboratory's reference frame, the potential energy flux is zero and the total energy flux is the sum of the  three previous ones. The remaining terms are the source terms.  As the total energy is conserved, these terms correspond to the conversions of one kind of energy into another. Thus, the work of the electric force involves an exchange of the electric potential and the kinetic energies, and the Joule effect and  the viscous dissipation involve the conversions of the potential and the kinetic energies into the internal energy, respectively.

Finally, the balance equation for the volume entropy $S$  can be written as 
\begin{eqnarray}
\rho \frac{D}{Dt}\left( \frac{S}{\rho }\right) =s-div\vec{\Sigma } ,
\end{eqnarray}
where $s$ and $\vec{\Sigma}$ represent the entropy volume production and the flux, respectively.

\subsection{Thermodynamic relations and entropy production}

We postulate the local thermodynamic equilibrium, namely, each R.E.V. is assumed to be in a state of thermodynamic equilibrium which is different from one point to another. Similarly as  for the balance equations, we derive the Gibbs \cite{deGroot}, Euler and Gibbs-Duhem relations of the material 
\begin{eqnarray}
T \frac{D}{Dt} \left( \frac{S}{\rho} \right)  \nonumber \\
=\frac{D}{Dt} \left( \frac{U}{\rho} \right) +p \frac{D}{Dt} \left( \frac{1}{\rho} \right)- \frac{1}{\rho}\utilde{\sigma^{e}}^{s}:grad\vec{V} ,\\
p=TS-U+\sum\limits_{k=1,2}\mu _{k}\rho _{k} ,\\
\phi _{1}grad p=S_{1}grad T+\rho _{1}grad \mu_{1} ,\\
\phi _{2}grad p=S_{2}grad T+\rho _{2}grad \mu _{2}- \sigma_{ij}^{es} grad \epsilon_{ij}^{s} ,
\end{eqnarray}
where $T$ represents the absolute temperature, $p$ denotes the pressure, $\mu_{k}$  stands for the mass chemical potential, and $\phi_{k} =\left\langle \chi_{k} \right\rangle$ is the volume fraction of phase $k$. $\utilde{\sigma^{e}}$ and $\utilde{\sigma^{v}}$ refer to the equilibrium  and dynamic (or viscous) stress tensor, and superscript $^{s}$ indicates the deviatoric part of a second-rank tensor 
\begin{eqnarray}
\utilde{\sigma}=-p\utilde{1}+\utilde{\sigma ^{e}}^{s}+\utilde{\sigma ^{v}} .
\end{eqnarray}
$\utilde{\epsilon}$ is the strain tensor and the strain rate tensor is :
\begin{eqnarray} 
\dot{\utilde{\epsilon}}=\frac{1}{2}\left( grad\vec{V}+grad\vec{V}^{T}\right) ,
\end{eqnarray}
where $\; \dot{ } \;$ denotes a time derivative.

By combining the Gibbs relation with the internal energy, and the mass balance equations, the entropy flux of the system, and the production can be  accordingly determined 
using identification with the entropy balance equation 
\begin{eqnarray}
\fl s=\frac{1}{T}\utilde{\sigma^{v}}:grad\vec{V}  +\frac{1}{T} \vec{E}.\vec{i}-\frac{1}{T^{2}} \vec{Q}.grad T  \nonumber\\
+\sum \limits_{1,2} \rho_{k} (\vec{V}-\vec{V_{k}}).grad \left( \frac{\mu _{k}}{T} \right) ,\\
\vec{\Sigma}=\frac{\vec{Q}}{T} +\sum \limits_{1,2} \left( \frac{\mu _{k} \rho_{k}}{T} + S_{k} \right) (\vec{V}-\vec{V_{k}}) ,                     
\end{eqnarray}
where the heat flux $\vec{Q}$ is defined by 
\begin{eqnarray}
\vec{Q} = \vec{Q'}+ \sum\limits_{1,2} \left[ U_{k}(\vec{V_{k}} - \vec{V}) -\utilde{\sigma _{k}}.(\vec{V_{k}}-\vec{V}) \right] .
\end{eqnarray}

Each term of entropy production corresponds to an irreversible process (the viscous friction, the Joule effect, the heat conduction, and the mass diffusion), and it is the product of a flux and the corresponding generalized force. The sum of the two mass fluxes $\rho_{k} (\vec{V}-\vec{V_{k}})$ is zero, indicating their linear dependency ; we choose to express the entropy production as a function of the diffusion flux of the cations in the solid 
\begin{eqnarray}
\vec{J_{m}}=\rho_{1} (\vec{V_{1}}-\vec{V_{2}}) .
\label{Jm}
\end{eqnarray}
We can identify a scalar flux, a second-order tensorial flux, and two linearly independent vectorial fluxes, as well as the corresponding generalized forces (\tref{tableF}).
\begin{table}[h!]
\caption{\label{tableF}Generalized forces and fluxes.} 
\begin{indented}
\lineup
\item[]\begin{tabular}{ll}
\br
Fluxes & Generalized forces\\ \mr
$\frac{1}{3}tr\left( \utilde{\sigma ^{v}}\right)
\quad $ & $\frac{1}{T}div\vec{V}$ \\
$\vec{Q}$ & $grad\left( \frac{1}{T}%
\right) $ \\
$\vec{J_{m}}$ & $\frac{\rho_{2}}{\rho} \left[ \frac{Z_{1}-Z_{2}}{T} \vec{E}+ grad \left( \frac{\mu_{2}-\mu_{1}}{T} \right) \right]$ \\
$\utilde{\sigma ^{v}}^{s}$ & $\frac{1}{T}%
grad\vec{V}^{s}$ \\ \br
\end{tabular}
\end{indented}
\end{table}

\subsection{Constitutive equations}

When the state of the system is close to equilibrium, the thermodynamics of linear irreversible processes allows us to establish three constitutive equations. We have assumed that the medium is isotropic. According to Curie's symmetry principle, there cannot be any coupling between the fluxes and the forces if their tensorial orders differ by one unit; moreover, couplings between the fluxes and the forces of different tensorial orders are typically negligible \cite{deGroot}.
We assume that at equilibrium, the material satisfies the Hooke's law 
\begin{eqnarray}
\utilde{\sigma^{e}}=\lambda \left( tr\utilde{\epsilon}\right) \utilde{1}+\frac{E}{1 + \nu}\utilde{\epsilon} ,
\label{sigma-e}
\end{eqnarray}
with 
\begin{eqnarray}
p = -\frac{1}{3} tr \utilde{ \sigma^{e}} = -\frac{E}{3(1-2 \nu)}tr \utilde{\epsilon} ,
\end{eqnarray}
where $\lambda$ denotes the first Lamé constant, $E$ the Young's modulus and $\nu$ the Poisson's ratio. We then obtain a Kelvin-Voigt type rheological equation 
\begin{eqnarray}
\utilde{\sigma}=\lambda \left( tr\utilde{\epsilon}\right) \utilde{1}+\frac{E}{1 + \nu}\utilde{\epsilon} +\lambda _{v}\left( tr \dot{\utilde{\epsilon}}\right) \utilde{1}+2\mu _{v}\dot{\utilde{\epsilon}} ,
\end{eqnarray}
where $\lambda _{v}$ and $\mu _{v}$ are two viscoelastic coefficients.
Since the medium is assumed to be isotropic, the two vectorial constitutive equations can be written in the following form 
\begin{eqnarray}
\vec{Q} = -\frac{L_{qq}}{T^{2}} grad T \nonumber \\
+  \frac{L_{qj} \rho_{2}}{\rho} \left[ \frac{Z_{1}-Z_{2}}{T} \vec{E}+ grad \left( \frac{\mu_{2}-\mu_{1}}{T} \right) \right] , \\
\vec{J_{m}}=-\frac{L_{jq}}{T^{2}} grad T \nonumber \\
+ \frac{L_{jj} \rho_{2}}{\rho} \left[ \frac{Z_{1}-Z_{2}}{T} \vec{E}+ grad \left( \frac{\mu_{2}-\mu_{1}}{T} \right) \right] ,
\end{eqnarray}
where the scalar phenomenological coefficients satisfy the Onsager reciprocal relation :
\begin{eqnarray}
L_{qj} = L_{jq} ,
\end{eqnarray}
and  
\begin{eqnarray}
L_{qq} > 0 \qquad L_{jj} > 0 \qquad L_{qj}^{2} \leq L_{qq} L_{jj}.
\end{eqnarray}
The first vectorial constitutive relation is a generalized form of Fourier's law. In the isothermal case and using equation (\ref{Jm}), the second relation takes the form of a generalized Darcy's law 
\begin{eqnarray}
\vec{V_{1}}-\vec{V_{2}}= -\frac{K}{\eta \phi_{1} } \left[ grad p  + \left( \frac{1} {\rho_{2}^{0}} - \frac{1}{\rho _{1}^{0}} \right)^{-1} \right. \nonumber \\
\left. \left( (Z_{1}-Z_{2})\vec{E} +\frac{1}{\phi_{2} \rho_{2}^{0}}\sigma_{ij}^{es} grad \epsilon_{ij}^{s} \right) \right] ,
\end{eqnarray}
where $K$ is the intrinsic permeability of the solid phase, and $\eta$ the dynamic viscosity of the liquid.

\section{Application to a cantilever blade}
In view of validating our model, we applied it to the case of a pseudo trilayer cantilever bending under the action of a permanent electric potential difference (static case). 
In addition, we assume that the evolution is isothermal.

\subsection{Modeling of the bending beam}
We chose a reference frame $Oxyz$ such that the axis $Ox$ is along the length of the undeformed beam, the axis $Oz$ is orthogonal to the beam and the axis $Oy$ is 
along its width (\fref{Forces}). We use the standard hypothesis of Bernoulli and Barré Saint Venant.
\begin{figure}[h]
\begin{center}
\includegraphics[height=2.7cm]{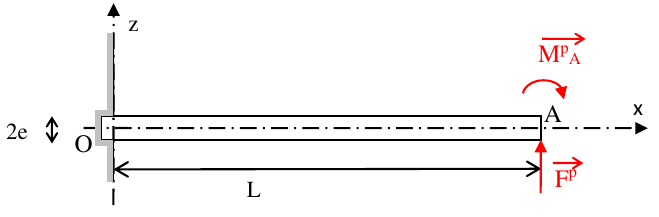}
\caption{Forces exerted on the beam.}
\label{Forces}
\end{center}
\end{figure}

The blade being studied is $2e = 250 \mu m$ thick, $2 \ell = 11 mm$ wide and $18 mm$ long. Since the blade is thin and the measurement point is located at $L=3 mm$ from the clamped end $O$, a beam model with small deformations and small displacements is sufficient. The end $A$ can either be free or subject to a shear force $\vec{F^{p}}$ blocking its displacement. When an electric potential difference $2 \varphi_{0} = 2~Volts$ is applied, the cations move towards the negative electrode and swell it, causing the blade 
to bend towards the positive electrode. This bending can be be modeled by a bending moment $\vec{M^{p}}$ around the axis $Oy$ axis 
\begin{eqnarray}
M^{p}= \int_{-l}^{l} \int_{-e}^{e}\sigma_{xx}~z~dz~dy = -6l \int_{-e}^{e} p~z~dz .
\label{moment}
\end{eqnarray}

\begin{wrapfigure}{r}{0.25\textwidth}
\includegraphics[width=\linewidth]{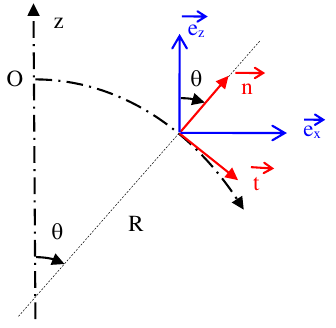}
\caption{Bending beam.}
\label{Poutre-flex}
\end{wrapfigure}

The beam undergoes pure bending. We derive the expression for the deflection $w$, the angle of rotation of the end section $\theta$ and the deformation $\epsilon_{xx}$ in the case of the cantilever beam, as well as the blocking force $\vec{F^{p}}$ (\fref{Poutre-flex}) 

\begin{eqnarray}
w=-\frac{M^{p}}{2E_{m}I^{p}}L^{2} ,\qquad \qquad & \theta = \frac{M^{p}}{E_{m}I^{p}}L ,\nonumber \\ 
\epsilon_{xx} =\frac{M^{p}e}{E_{m}I^{p}} ,&F^{p}=\frac{3}{2}\frac{M^{p}}{L} , \qquad \quad
\label{eq-meca}
\end{eqnarray}
where $I^{p}=\frac{4 \ell e^{3}}{3}$ denotes the moment of inertia with respect to the $Oy$ axis and $E_{m}$ the mean Young's modulus of the pseudo trilayer.

\subsection{Equations in the static case}

We consider a two-dimensional problem in the $Oxz$ plane. We assume that the axial components $E_{x}$ and $D_{x}$ of the local electric field and displacement are negligible compared to their respective transverse components $E_{z}$ and $D_{z}$. It is also assumed that the electric field $E_{z}$, the induction $D_{z}$, the potential $\varphi$ and 
the electric charge $\rho Z$, the pressure $p$ and the volume fraction of the cations $\phi_{1}$ depend only on $z$. With these assumptions, we have 
\begin{eqnarray}
p = -\frac{\sigma_{xx}}{3} ,
\end{eqnarray}
and
\begin{eqnarray}
\sigma_{ij}^{es} grad \epsilon_{ij}^{s} = \frac{6(1+ \nu)}{E}p \frac{dp}{dz} \vec{e_{z}} .
\end{eqnarray}
In addition, the trace of the strain tensor is equal to the relative variation of the volume of the material. Assuming that the solid phase is incompressible, 
we derive using (\ref{sigma-e}) 
\begin{eqnarray}
tr \utilde{\epsilon} =\frac{\phi_{1}-\phi_{1r}}{1-\phi_{1}}= -\frac{3 (1 - 2 \nu)}{E} p ,
\end{eqnarray}
where $\phi_{1r}$ is the volume fraction of the cations in the undeformed beam. The equations governing the system are reformulated as 
\begin{equation}
\cases{
E_{z}=-\frac{d \varphi}{dz} , \\
\frac{d D_{z}}{dz}= \rho Z=\phi _{1}(\rho _{1}^{0} Z_{1} -\rho_{2}^{0} Z_{2})+\rho_{2}^{0} Z_{2} , \\
D_{z}=\varepsilon E_{z} , \\
\frac{\rho _{1}^{0}-\rho_{2}^{0}}{\rho _{1}^{0} \rho_{2}^{0}}\frac{dp}{dz} + (Z_{1}-Z_{2})E_{z} +\frac{6(1+\nu)}{\phi_{2}\rho_{2}^{0}E} p \frac{dp}{dz}=0 , \\
\frac{\phi_{1}-\phi_{1r}}{1-\phi_{1}}= -\frac{3 (1 - 2 \nu)}{E} p .}
\end{equation}
When the beam is not deformed, we can write 
\begin{eqnarray}
\rho Z=\phi_{1r}(\rho_{1}^{0} Z_{1} -\rho_{2}^{0} Z_{2})+\rho_{2}^{0} Z_{2}=0 .
\end{eqnarray}
Moreover, the fourth equation of the system can be integrated for each layer accounting for the first and the fifth equations 
\begin{eqnarray}
\frac{\rho _{1}^{0}-\rho_{2}^{0}}{\rho _{1}^{0} \rho_{2}^{0}}p - (Z_{1}-Z_{2}) \varphi \nonumber \\
\qquad +\frac{3(1+\nu)}{\phi_{2r}\rho_{2}^{0}E} \left( p^{2}-\frac{2(1-2 \nu)}{E}p^{3} \right) =Cte .
\label{LD}
\end{eqnarray}
The system takes on the form  
\begin{equation}
\cases{
\overline{E}=-\overline{\varphi}' , \\
\overline{D}' =\overline{\rho Z}=\frac{\phi_{1}-\phi_{1r}}{1-\phi_{1r}}=- A_{0} \overline{\varphi}'' ,\\
\overline{D}=-A_{0} \overline{\varphi}' ,\\
\overline{p} - A_{2} \overline{\varphi} +\frac{A_{3}}{2 \phi_{2r}}  \left( \overline{p}^{2} -\frac{2 E_{m}}{3E}\overline{p}^{3} \right)=Cte , \\
\overline{p} =-\frac{E}{E_{m}}\frac{\phi_{1}-\phi_{1r}}{1-\phi_{1}} ,}
\label{Syst-eq}
\end{equation}
where the dimensionless variables are defined by 
\begin{eqnarray}
\overline{z}=\frac{z}{e} , &\overline{\varphi}=\frac{\varphi}{\varphi_{0}} , &\overline{E}=\frac{e}{\varphi_{0}} E_{z} ,\nonumber\\
\overline{p}=\frac{3 ( 1-2 \nu)}{E_{m}} p ,\quad &\overline{\rho Z}=\frac{\rho Z}{\rho _{1}^{0}Z_{1}} , \quad &  \overline{D}=\frac{D_{z}}{\rho _{1}^{0}Z_{1}e} ,
\end{eqnarray}
and the dimensionless constants by 
\begin{eqnarray}
A_{0} = \frac{\varepsilon \varphi_{0}}{\rho _{1}^{0}Z_{1}e^{2}} , \qquad & A_{2}= \frac{3(1-2 \nu) \varphi_{0} \rho_{2}^{0}(Z_{1}-Z_{2})}{E_{m}\left(1-\frac{\rho_{2}^{0}}{\rho_{1}^{0}}\right)} ,\nonumber \\
A_{1}=A_{0} \frac{E}{E_{m}} , &
A_{3}=  \frac{2(1+ \nu) E_{m}}{(1-2 \nu) E \left( 1-\frac{\rho_{2}^{0}}{\rho_{1}^{0}} \right)} .
\label{Const}
\end{eqnarray}
$'$ indicates a derivative with respect to $\overline{z}$. The boundary conditions and the electroneutrality are written as 
\begin{eqnarray}
\overline{\varphi}(-1) = 1 , \qquad \qquad \overline{\varphi}(1) = -1
\label{CL} \\
\int_{-1}^{+1} \overline{\rho Z} d\overline{z} =0 .
\end{eqnarray}
Assuming that the permittivity is constant throughout the blade, hypothesis which provided the best results for ionic polymers \cite{Tixier4}, the latter condition offers 
\begin{eqnarray}
\overline{\varphi}'(-1) =\overline{\varphi}'(1) .
\label{electroneut}
\end{eqnarray}

\begin{figure}[h]
\begin{center}
\includegraphics[height=3cm]{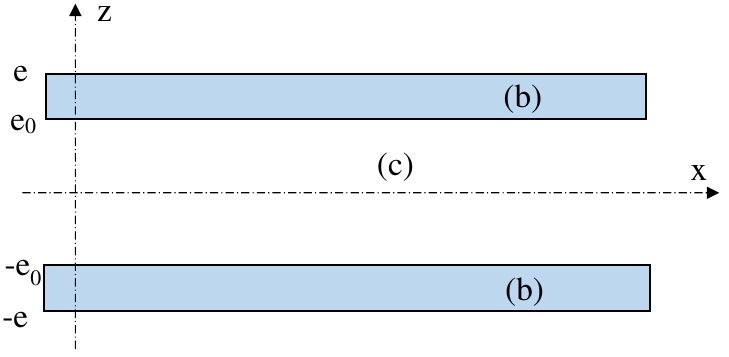}
\caption{Pseudo trilayer.}
\label{Tricouche}          
\end{center}
\end{figure}

In addition, the pressure, the electric potential and its first derivative must be continuous through the interfaces between two layers, since there is no accumulation of electric 
charge on them. The subscripts $_{b}$ and $_{c}$ refer to the electrodes and center respectively (\fref{Tricouche}). Denoting $\pm \overline{e_{0}}$ the z-coordinates of the interfaces,  the interfaces conditions are read as 

\begin{eqnarray}
\overline{p_{b}}(\pm \overline{e_{0}}) = \overline{p_{c}}(\pm \overline{e_{0}}) , \nonumber \\
\overline{\varphi_{b}}(\pm \overline{e_{0}}) = \overline{\varphi_{c}}(\pm \overline{e_{0}}) , \nonumber \\
\overline{\varphi_{b}}'(\pm \overline{e_{0}}) = \overline{\varphi_{c}}'(\pm \overline{e_{0}}) .
\label{cont}
\end{eqnarray}

\subsection{Solving the beam equation}

The density and the electrical mass charge of the liquid phase are  $\rho_{1}^{0}= 1.53g~cm^{-3}$ and $Z_{1}=8.69~10^{5}kg~m^{-3}$, respectively, and the Poisson's ratio can be estimated at $\nu \simeq 0.4$, which is a typical value for polymers \cite{FannirTh}. The concentration of PEDOT varies throughout the thickness; as a result, the blade can be assimilated to a pseudo trilayer : the two outer layers or electrodes, about $e_{b}=30 \mu m$ thick, are rich in PEDOT, while being almost devoid of EMITFSI at rest, 
the central part which is $2 e_{c}=190\mu m$ thick, behaves like an ion reservoir. 
The other quantities can be approximated as rectangular functions. To recompile them, we use the thickness of the dry blade ($142 \mu m$ ; courtesy of C. Plesse) and its estimated density ($1.01g~cm^{-3}$), as well as the molar masses of EMITFSI ($391.21g~mol^{-1}$) and $TFSI^{-}$ anions ($280.15g~mol^{-1}$). The absorbed mass of EMITFSI is approximately a decreasing affine function of the PEDOT mass when the PEDOT mass fraction is low \cite{Festin}. We derive the mass fractions of EMITFSI in each layer, as well as the density, the mass electric charge and the volume fraction of the solid phase  at the rest (\tref{Table-donnees}).

The Young's modulus varies exponentially with the mass fraction of PEDOT \cite{FannirTh}. The mean tensile Young's modulus of the blade is around $E_{tr}\simeq 30 MPa$, which  aligns closely with the measurements of \citeasnoun{Festin2013} and \citeasnoun{Woehling2018}. For the central part, we estimate it to be around $E_{c}=15 MPa$. We can 
determine the Young's modulus of the electrodes using a mixing law 
\begin{eqnarray}
E_{b} = \frac{E_{tr} e - E_{c}e_{c}}{e_{b}} \simeq 70 MPa ,
\end{eqnarray}
which is confirmed by the  measurements of \citeasnoun{Woehling2018}. The average Young's modulus can be calculated from the aforementioned values or from measurements of the deflection and the blocking force \cite{festin2014} 
\begin{eqnarray}
E_{m}=E_{c}\frac{e_{c}^{3}}{e^{3}}+E_{b} \left( 1-\frac{e_{c}^{3}}{e^{3}} \right)=\frac{L^{3}}{4l e^{3}}\frac{F}{w}\simeq 40 MPa .
\end{eqnarray}

\renewcommand{\arraystretch}{2}
\begin{table*}[t!]
\caption{\label{Table-donnees}Values of the quantities in each layer.} 
\begin{indented}
\lineup
\item[]\begin{tabular}{ccccccc}
\br
&PEDOT&EMITFSI&$\rho_{2}^{0}$&$Z_{2}$&$\phi_{2r}$&$E$ \\ \mr
&$wt \%$&$wt \%$&$g~cm^{-3}$&$kg~m^{-3}$&$\%$ &$MPa$\\
Electrodes ($b$) &$26.2$&$33.9$&$1.11$&$-9.25~10^4$&$92.3$&$70$ \\
Center ($c$) &$2.58$&$67.7$&$1.27$&$-2.07~10^{5}$&$83.5$&$15$ \\ \br
\end{tabular}
\end{indented}
\end{table*}

The mean pressure can be estimated by measuring the blocking force $F^{p} \simeq 30 mN$ \cite{festin2014} 
\begin{eqnarray}
|p| \simeq \frac{\sigma_{xx}}{3} = \frac{E_{m} \epsilon_{xx}}{3} = \frac{L}{6 \ell e^{2}}F^{p} .
\end{eqnarray}
We deduce that the last two terms of the equation (\ref{LD}) represent less than $5\%$ of the first one, and therefore they can be neglected 
\begin{eqnarray}
\overline{p} - A_{2} \overline{\varphi} =Cte .
\label{DarcySimp}
\end{eqnarray}

The relative volume variations of the blade are approximately $tr \utilde{\epsilon} \simeq \epsilon_{xx} \sim 1.4\%$ \cite{Festin,festin2014}. To simplify solution, we first 
neglect them, allowing us to write the last equation of (\ref{Syst-eq}) 
\begin{eqnarray}
\overline{p}=-\frac{E}{E_{m}}(\phi_{1}-\phi_{1r})=A_{1} \phi_{2r} \overline{\varphi}''
\end{eqnarray}
Combining it with the penultimate equation in (\ref{Syst-eq}) we obtain 
\begin{eqnarray}
\overline{\varphi}''-\delta^{2} \overline{\varphi} = -\delta^{2} B_{2}
\label{EDapp}
\end{eqnarray}
where $B_{2}$ is an unknown constant in each layer and 
\begin{eqnarray}
\delta = \sqrt{3 (1-2 \nu) \frac{\rho _{1}^{0}Z_{1}e^{2}}{\varepsilon}\frac{\rho_{2}^{0}(Z_{1}-Z_{2})}{E\phi_{2r} \left(1-\frac{\rho_{2}^{0}}{\rho_{1}^{0}}\right)}} .
\end{eqnarray}
It can be analytically integrated in each layer, accounting for the conditions (\ref{CL}), (\ref{electroneut}) and (\ref{cont}) 
\begin{eqnarray}
\overline{\varphi} =-\frac{sh(\delta \overline{z})}{sh \delta} .
\end{eqnarray}
In this scenario, we deduce that $B_{2}$  is approximately equal to zero in all layers.
By combining the second and last equation of (\ref{Syst-eq}) with (\ref{DarcySimp}), we obtain a differential equation that governs the electric potential leading 
to more accurate results 
\begin{eqnarray}
\overline{\varphi}''=\frac{1}{A_{0}} \left[ -1 +\frac{1}{1-A_{4}(\overline{\varphi }- B_{1})} \right] ,
\label{ED}
\end{eqnarray}
where $B_{1}$ is an unknown constant a priori different in each layer and 
\begin{eqnarray}
A_{4} =\frac{A_{0}A_{2}}{A_{1}} .
\end{eqnarray}
Given the results obtained with the analytical solution, the blade behaves as a conductor. We deduce that $B_{1}$ has the same value throughout the entire blade. 
The condition of electroneutrality then leads to 
\begin{eqnarray}
B_{1}=1-\frac{1}{A_{4b}}+\frac{2 e^{-2A_{4b}}}{1-e^{-2A_{4b}}} \simeq 1-\frac{1}{A_{4b}} .
\end{eqnarray}
(\ref{ED}) can be numerically integrated given the boundary conditions (\ref{CL}), (\ref{electroneut}) and (\ref{cont}). The pressure satisfies 
\begin{eqnarray}
\overline{p}=A_{2}(\overline{\varphi}-B_{1}) .
\label{pres}
\end{eqnarray}
The values of the miscellaneous quantities at the surface of the electrodes are listed in Table \ref{Bound-val}, and the constants are detailed in Table \ref{Constantes}.
\renewcommand{\arraystretch}{2}
\begin{table*}[t!]
\caption{\label{Bound-val}Boundaries values.} 
\begin{indented}
\lineup
\item[]\begin{tabular}{cccccc}
\br
&$\overline{\varphi}$&$\overline{D}$&$\overline{\rho Z}$&$\overline{p}$&$\phi_{1}$ \\ \mr
$\overline{z}=-1$&$1$&$\sqrt{\frac{2 A_{0}}{A_{4b}}(2 A_{4b}-1-\ln2-\ln A_{4b})}$&$-\frac{e^{2 A_{4b}}}{2 A_{4b}}$&$\frac{A_{2b}}{A_{4b}}$&$-\frac{\phi_{2rb}}{2 A_{4b}}e^{2 A_{4b}}$ \\
Center&$0$&$0$&$0$&$0$&$1-\phi_{2rc}$ \\
$\overline{z}=1$&$-1$&$\sqrt{\frac{2 A_{0}}{A_{4b}}(2 A_{4b}-1-\ln 2-\ln A_{4b})}$&$1-\frac{1}{2 A_{4b}}$&$\frac{A_{2b}}{A_{4b}} (1-2A_{4b}) $&$1-\frac{\phi_{2rb}}{2 A_{4b}}$ \\ \br
\end{tabular}
\end{indented}
\end{table*}

\renewcommand{\arraystretch}{2}
\begin{table*}[t!]
\caption{\label{Constantes}Values of the dimensionless constants of the model (relations \ref{Const})}. 
\begin{indented}
\lineup
\item[]\begin{tabular}{ccccccccc}
\br
 & $ A_{0}$ &$ A_{1}$ &$A_{2}$ & $A_{3}$&$ A_{4}$ &$\delta$&$B_{1}$&$B_{2}$ \\ \mr
Electrodes&$4.81~10^{-2} \varepsilon$ & $8.43~10^{-2} \varepsilon$ &$58.4$ & $29.2$&$33.4$ &$27.3/\sqrt{\varepsilon}$&$0.970$&$0$\\ 
Center&$4.81~10^{-2} \varepsilon$ & $1.81~10^{-2} \varepsilon$ &$120$ &$218$&$319$&$89.1/\sqrt{\varepsilon}$&$0.970$&$0$ \\ \br
\end{tabular}
\end{indented}
\end{table*}

\section{Numerical results and discussion}

The static dielectric permittivity of the pseudo trilayer has not been measured. However, the permittivity of PEDOT is estimated at $8~10^{-6} Fm^{-1}$ at $10Hz$ \cite{Taj2020}. Meanwhile \citeasnoun{Ninis} measured a permittivity of $3~10^{-8}~Fm^{-1}$ for the PEDOT/poly(n-vinylcarbazole) copolymer at $50~Hz$. The permittivity of NBR containing $44\%$ acrylonitrile is approximately $10^{-8} Fm^{-1}$ \cite{Vennemann}. PEO saturated with aqueous $LiClO_{4}$ solution (another liquid used to make pseudo trilayers giving similar results) has a permittivity close to $10^{-6} Fm^{-1}$ \cite{Das,Karmakar}. Moreover, the permittivity of the blade is a priori between $10^{-8} Fm^{-1}$ and $10^{-6} Fm^{-1}$. By comparison, that of the Nafion was estimated at $5~10^{-7} Fm^{-1}$ \cite{Deng,Tixier4}.

Permittivity is the only adjustable parameter of our model. We adjust it to closely match the deflection and the blocking force values published in the literature. Consequently we obtain a permittivity equal to $2~10^{-7} Fm^{-1}$ for the analytical model and $10^{-8} Fm^{-1}$ for the numerical simulations. As the numerical model is more accurate than the analytical one, we can assume that the permittivity of the blade is about $10^{-8} Fm^{-1}$, which aligns with the previous estimations.

\subsection{Deflection and blocking force}

For these permittivity values, we evaluated the blocking force $F^{p}$, the deflection $w$, the deformation $\epsilon_{xx}$ and the angle of rotation of the end section $\theta$ at $L=3mm$ from the clamp. \Tref{Mes} provides a summary of the values obtained from both the analytical and the numerical models, along with the available experimental data. 
Our results show good agreement with the measurements reported by \citeasnoun{festin2014}.

\begin{table}[t!]
\caption{\label{Mes}Comparison between experimental mechanical quantities from Festin et al. \cite{Festin,festin2014} and the results of our model ; the experimental deflection and, the angle of rotation are calculated using the relations (\ref{eq-meca}).} 
\begin{indented}
\lineup
\item[]\begin{tabular}{@{}cccc}
\br
&Experiment&Analytical model&Numerical model\\ \mr
$w~(mm)$&$0,5$&$0,51$&$0,48$\\
$F^{p}~(mN)$&$30$&$32.9$&$30.6$\\
$\epsilon_{xx}~(\%)$&$1.4$&$1.43$&$1,33$\\
$\theta~(deg)$&$19.1$&$19.7$&$18.3$\\ \br
\end{tabular}
\end{indented}
\end{table}

\subsection{Scaling laws}
The bending moment can be calculated using the analytical solution 
\begin{eqnarray}
M = 4 l e \varphi_{0} \sqrt{\frac{3 \varepsilon E \phi_{2r}}{1-2 \nu} \frac{\rho_{2}^{0}}{\rho_{1}^{0}-\rho_{2}^{0}} \frac{Z_{1}-Z_{2}}{Z_{1}}} .
\end{eqnarray}
We derive the scaling law for the deflection, the blocking force, the deformation and the angle of rotation, which are identical to those obtained for the Nafion \cite{Tixier4} 
\begin{eqnarray}
w \sim \frac{L^{2} \varphi_{0}}{e^{2}} ,\qquad \qquad &\theta  \sim\frac{L \varphi_{0}}{e^{2}} , \nonumber \\ 
\epsilon_{xx} \sim \frac{\varphi_{0}}{e} ,& F^{p} \sim \frac{l e \varphi_{0}}{L} ,
\end{eqnarray}
These scaling laws agree well with the published experimental results : especially, the deformation and the blocking force are proportional to the imposed potential (\citeasnoun{Festin2013}, \citeasnoun{nguyen2018} and \citeasnoun {Alici2007} for a similar three-layer material) ; the blocking force is inversely proportional to the length, and the deflection  increases with the length \cite{Woehling}. Lastly, for the PPy-based trilayers, the blocking force increases almost linearly with the width \cite{Alici2007}.

\subsection{Profiles of the mechanical quantities}

Figures \ref{phi}, \ref{D}, \ref{rhoZ} and \ref{p} show the profiles of the electric potential, the displacement, and the electric charge as well as the pressure in the blade thickness 
for both the analytical and numerical models. The curves exhibit steep variations near the boundaries and remain relatively constant in the center. Notably, the electric displacement and the electric charge are zero, while the electric potential remains constant throughout most of the interval, resembling the behavior of a conductor. The characteristic length over which these quantities vary is about $10$ nanometres.

The numerical model displays dissymmetrical profiles with respect to the blade center, which is more realistic than the analytical model showing the asymmetrical displacement of the ions.

The electric charge profile shows a small plateau approximately  $10nm$ near of the upper electrode corresponding to the region where the cations accumulate. Simultaneously, 
the lower electrode becomes highly electronegative, indicating that the cations of this region have migrated towards the central part.

No experimental curves are available for these profiles. However, they match with those from various models, particularly for the Nafion, specially for the electric charge, the electric potential, and the electric field \cite{nemat2002,wallmersperger2009,nardinocchi2011}. In addition, the deflection and the blocking force are derived from the pressure profile and their values, as well as their variations with the imposed potential, the blade length, and the thickness of the blade, closely match the experimental measurements.

\begin{figure*}[h!]
\includegraphics[width=1.0\textwidth]{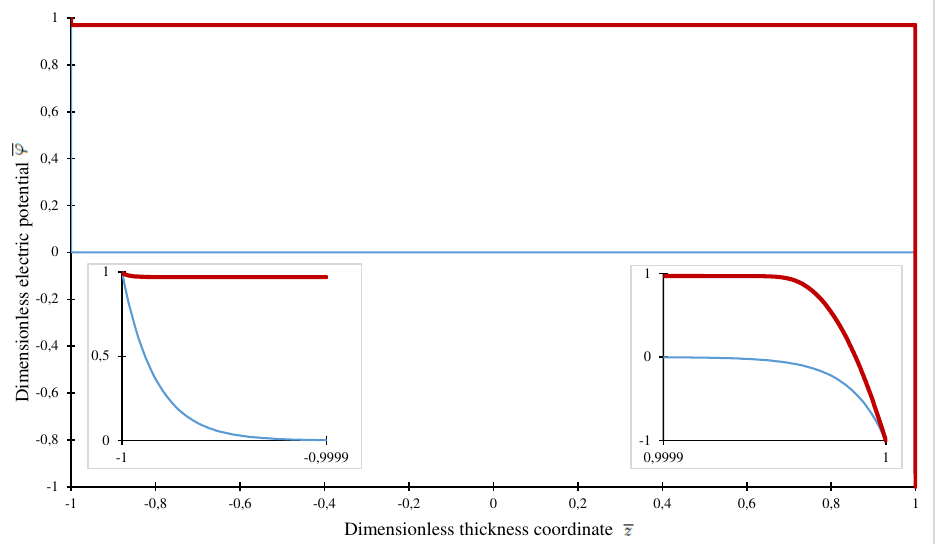}
\caption{Variation of the dimensionless electric potential $\overline{\varphi}=\frac{\varphi}{\varphi_{0}}$ in the thickness $\overline{z}=\frac{z}{e}$ of the blade ; the distribution at the vicinity of the boundaries are detailled in insets. The analytical model is in thin blue line, the numerical simulation in thick red line.}
\label{phi}
\end{figure*}

\begin{figure*}[h!]
\includegraphics[width=1.0\textwidth]{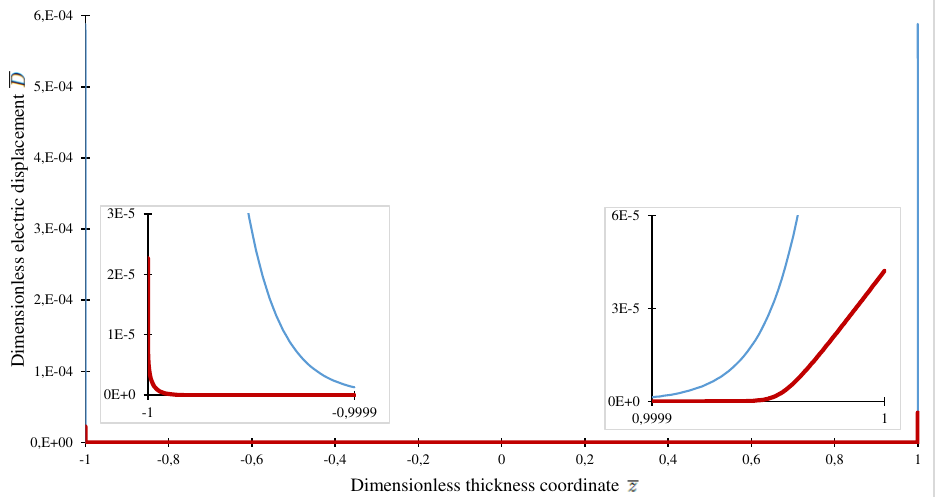}
\caption{Variation of the dimensionless electric displacement $\overline{D}=\frac{D_{z}}{\rho _{1}^{0}Z_{1}e}$ in the thickness $\overline{z}=\frac{z}{e}$ of the blade ; the distribution at the vicinity of the boundaries are detailled in insets (same colors as in \fref{phi})}.
\label{D}
\end{figure*}

\begin{figure*}[h!]
\includegraphics[width=1.0\textwidth]{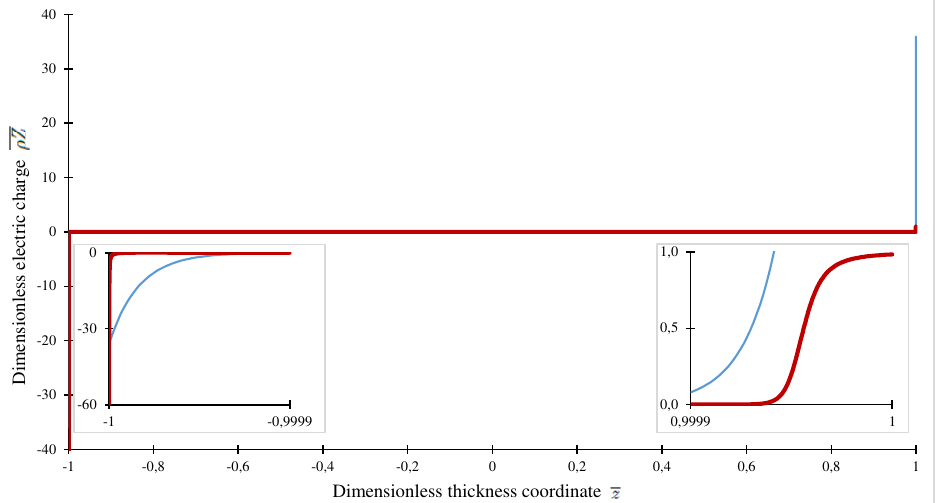}
\caption{Variation of the dimensionless electric charge density $\overline{\rho Z}=\frac{\rho Z}{\rho _{1}^{0}Z_{1}}$ in the thickness $\overline{z}=\frac{z}{e}$ of the blade ; the distribution at the vicinity of the boundaries are detailled in insets (same colors as in \fref{phi}).}
\label{rhoZ}
\end{figure*}

\begin{figure*}[h!]
\includegraphics[width=1.0\textwidth]{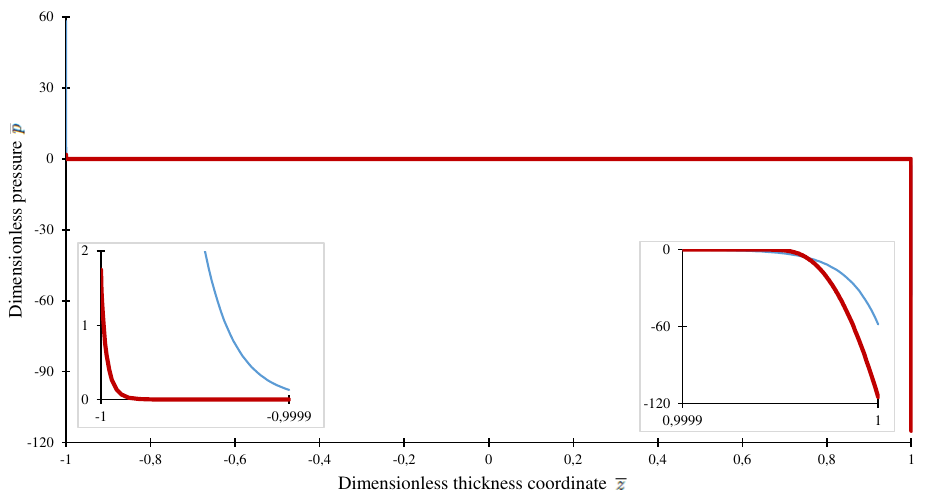}
\caption{Variation of the dimensionless pressure $\overline{p}=\frac{3 ( 1-2 \nu)}{E_{m}} p$ in the thickness $\overline{z}=\frac{z}{e}$ of the blade ; the distribution at the vicinity of the boundaries are detailled in insets (same colors as in \fref{phi}).}
\label{p}
\end{figure*}

\section{Conclusion}

In our research, we examined a pseudo trilayer consisting of three interpenetrating polymers saturated with an ionic liquid. One of these polymers, PEDOT, is an electro-active semiconductor polymer. We established the balance equations for this system. Its constitutive relations were rigorously deduced using the thermodynamics of linear irreversible processes. It is worthwhile noting that the present model does not rely on any empirical laws. The only adjustable parameter is the permittivity of the blade, which 
has not been experimentally measured but it is compatible with the data available in the literature. Unlike "black box" models, the present model allows us to determine the profiles of the different quantities inside the blade and can be easily adapted to similar materials, as demonstrated with the Nafion \cite{Tixier4}.

We then have applied the model to analyze a blade clamped at one end bending under the action of an electric potential difference between its two faces in the static and isothermal case ; the other end is either free or subject to a blocking force. We plotted the profiles of the quantities describing the blade such as the electric potential, the induction, and the 
electric charge as well as the pressure. The curves obtained, which are mainly constant in the central part of the blade thickness and vary very sharply near the boundaries, 
indicate that the material behaves like a conductor. Furthermore, we estimated the values of the strain and of the blocking force, which align well with the experimental data 
available in the literature.

Our next step, by means of the present model, is to study the dynamic case and the inverse effect.

\ack
We thanks C. Plesse and G.T.M. Nguyen for fruitful discussions and for their experimental data.

\section*{Main notations}

$k=1,2,i$ subscripts represent the cations, the solid ,and the interface, respectively. Quantities without subscripts refer to the whole material. Superscript $^{0}$ denotes a local quantity ; the lack of superscript indicates average quantity at the macroscopic scale. $\otimes$ denotes the dyadic product. Superscript $^{s}$ indicates the deviatoric part of a second-rank tensor, and $^{T}$ its transpose. $\; \dot{ } \;$ is a time derivative. Subscript $_{r}$ refers to the undeformed state, $_{b}$ to the outer layers and $_{c}$ to the central part. Over-lined letters represent dimensionless quantities.

\begin{description}
\item $A_{i}$, $B_{i}$ : dimensionless constants ;

\item $\vec{D}$ : electric displacement field ;

\item $e$ : half-thickness of the blade ;

\item $E$ ($E_{m}$, $E_{tr}$) : Young's modulus (mean bending and tensile Young's modulus) ;

\item $E_{c}$, $E_{p}$, $E_{tot}$ : kinetic, potential and total energies ;

\item $\vec{E}$ : electric field ;

\item $\vec{F^{p}}$ : blocking force ;

\item $\vec{i}$ : diffusion current ;

\item $\vec{I}$ : current density vector ;

\item $\vec{J_{m}}$ : diffusion flux of the cations in the solid ;

\item $I^{p}$ : moment of inertia with respect to the $0y$ axis ;

\item $K$ : intrinsic permeability of the solid phase ;

\item $l$ : half-width of the blade ;

\item $L$ : length of the blade ;

\item $\vec{M^{p}}$ : bending moment ;

\item $\vec{n_{k}}$ : outward-pointing unit normal of phase $k$ ;

\item $p$ : pressure ;

\item $\vec{Q}$, $\vec{Q'}$ : heat flux, conduction heat flux ;

\item $s$ : rate of entropy production ;

\item $S$ ($S_{k}$) : entropy density (of phase $k$) ;

\item $T$ : absolute temperature ;

\item $U$ ($U_{k}$) : internal energy density (of phase $k$) ;

\item $\vec{V}$ ($\vec{V_{k}}$, $\vec{V_{k}^{0}}$) : velocity (of phase $k$) ;

\item $w$ : deflection of the beam ;

\item $Z$ ($Z_{k}$) : total electric charge per unit of mass (of phase $k$) ;

\item $\varepsilon$ : absolute permittivity ;

\item $\utilde{\epsilon}$ : strain tensor ;

\item $\eta$ : dynamic viscosity of the liquid phase ;

\item $\theta$ : angle of rotation of the beam end section ;

\item $\lambda$ : first Lamé constant ;

\item $\lambda_{v}$, $\mu_{v}$ : viscoelastic coefficients ;

\item $\mu_{k}$ : mass chemical potential of phase $k$ ;

\item $\nu$ : Poisson's ratio ;

\item $\rho $ ($\rho^{0}_{k}$) : mass density (of phase $k$) ;

\item $\rho_{k}$ : mass concentration of phase $k$ ;

\item $\utilde{\sigma}$ ($\utilde{\sigma_{k}}$), $\utilde{\sigma^{e}}$, $\utilde{\sigma^{v}}$ : total (of phase $k$), equilibrium, dynamic stress tensors ;

\item $\vec{\Sigma}$ : entropy flux vector ;

\item $\phi_{k}$ : volume fraction of phase $k$ ;

\item $\varphi$ ($\varphi_{0}$) : electric potential (imposed electric potential) ;

\item $\chi_{k}$ : function of presence of phase $k$ ;

\end{description}

\section*{References}
\bibliographystyle{jphysicsB}
\bibliography{biblio_2}

\end{document}